\documentclass[11pt,twoside]{article}  
\usepackage{adassconf}
\usepackage{graphicx}

\begin{document}   

%
%
%

\paperID{P1.3.8}

%
%
%
%

\title{Grist: Grid-based Data Mining for Astronomy}

%
%
%

\author{Joseph C. Jacob, Daniel S. Katz, Craig D. Miller, Harshpreet Walia}
\affil{Jet Propulsion Laboratory, California Institute of Technology, Pasadena, CA 91109-8099}

\author{Roy Williams, S. George Djorgovski, Matthew J. Graham, Ashish Mahabal}
\affil{California Institute of Technology, Pasadena, CA 91125}

\author{Jogesh Babu, Daniel E. Vanden Berk}
\affil{The Pennsylvania State University, University Park, PA, 16802}

\author{Robert Nichol}
\affil{ICG, University of Portsmouth, PO1 2EG, UK}

%
%

\contact{Joseph Jacob}
\email{Joseph.Jacob@jpl.nasa.gov}

%
%
%
%
%

\paindex{Jacob, J. C.}
\aindex{Williams, R.}
\aindex{Babu, J.}
\aindex{Djorgovski, S. G.}
\aindex{Graham, M. J.}
\aindex{Katz, D. S.}
\aindex{Mahabal, A.}
\aindex{Miller, C. D.}
\aindex{Nichol, R.}
\aindex{Vanden Berk, D. E.}
\aindex{Walia, H.}

%
%

\authormark{Jacob, Williams, et al.}

%
%

\keywords{Grid services, Data mining, Astronomy, Virtual observatory,
Quasar search, Transient search}


\begin{abstract}          
The Grist project is developing a grid-technology based system
as a research environment for astronomy with massive and
complex datasets.  This knowledge extraction system
will consist of a library of distributed grid services controlled by a
workflow system, compliant with standards emerging from the grid
computing, web services, and virtual observatory communities.  This
new technology is being used to find high redshift quasars,
study peculiar variable objects, search for transients in
real time, and fit SDSS QSO spectra to measure black hole masses.
Grist services are also a component of the
``hyperatlas'' project to serve high-resolution multi-wavelength
imagery over the Internet.  In support of these science and outreach 
objectives, 
the Grist framework will provide the enabling fabric to tie together
distributed grid services in the areas of data access, federation,
mining, subsetting, source extraction, image mosaicking,
statistics, and visualization.
\end{abstract}

%
%

\section{Overview}

The Grist\footnote{Part of this research was carried out at the Jet Propulsion
Laboratory, California Institute of Technology, and was sponsored by
the National Science Foundation through an agreement with the National
Aeronautics and Space Administration.} project 
(\htmladdURL{http://grist.caltech.edu/}) is enabling 
astronomers and the public to interact with the grid
projects that are being constructed worldwide, and bring to flower the
promise of easy, powerful, distributed computing. Our objectives are
to understand the role of service-oriented architectures in
astronomical research, to bring the astronomical community to the grid
-- particularly TeraGrid, -- and to work with the National Virtual
Observatory (NVO) to build a library of compute-based web services.

The scientific motivation for Grist derives from creation and mining
of wide-area federated images, catalogs, and spectra. An astronomical
image collection may include multiple pixel layers covering the same
region on the sky, with each layer representing a different waveband,
time, instrument, observing condition, etc. The data analysis should
combine these multiple observations into a unified
understanding of the physical processes in the Universe. The familiar
way to do this is to
cross-match source lists extracted from different images. However, there is
growing interest in another method of federating images that
reprojects each image to a common set of pixel planes, then stacks
images and detects sources therein. While this has been done for
years for small pointing fields, we are using the TeraGrid to perform
this processing over wide areas of the sky in a systematic way, using
\htmladdnormallinkfoot{Palomar-Quest}{http://www.astro.caltech.edu/pq/} 
(PQ) survey data.  We expect this ``hyperatlas'' approach 
will enable us
to identify much fainter sources than can be detected in any individual
image; to detect unusual objects such as transients; and to deeply
compare (e.g., using principal component analysis) the large surveys
such as SDSS, 2MASS, DPOSS, etc. (Williams et al. 2003).

Grist is helping to build an image-federation pipeline for the
Palomar-Quest synoptic sky survey (Djorgovski et al. 2004), 
with the objectives of mining PQ data
to find high redshift quasars, to study peculiar variable
objects, and to search for transients in real-time (Mahabal et al. 2004).
Our PQ processing pipeline will use the TeraGrid for processing and will
comply with widely-accepted data formats and 
protocols supported by the VO community.

\section{Service-Oriented Architectures for Astronomy}
\label{sec:workflow}

\begin{figure}[hbt] 
\plotone{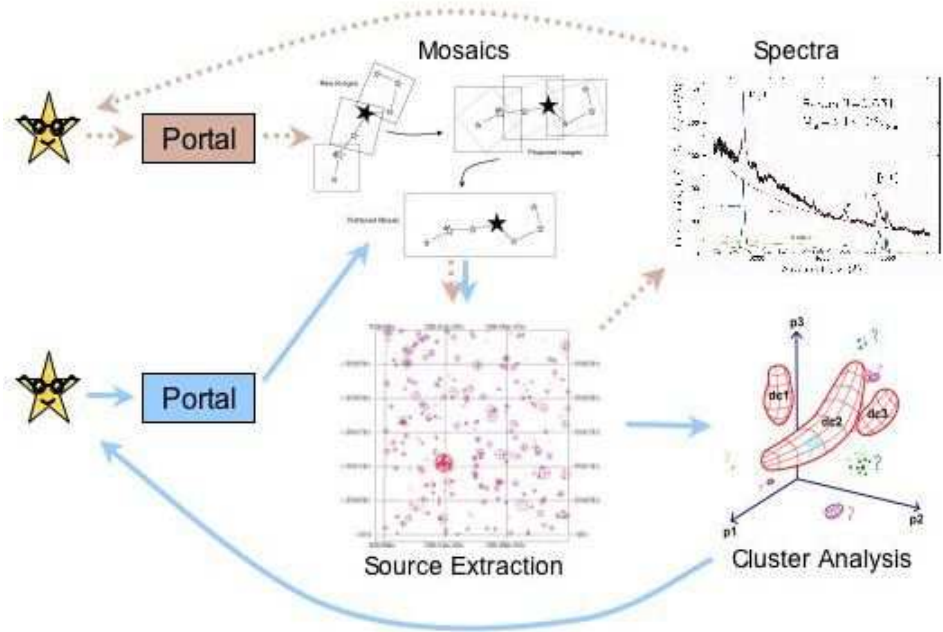}
\caption{Grist will deploy a library of interoperable services, which
may be composed in different ways for astronomical data mining
(e.g., two distinct workflows are indicated by the solid and dashed
arrows).}
\label{fig:workflow} 
\end{figure}

The Grist project is building web and grid services as well as the
enabling workflow fabric to tie together these distributed services in
the areas of data federation, mining, source extraction, image
mosaicking, coordinate transformations, data subsetting, statistics
-- histograms, kernel density estimation, and R language utilities
exposed by \htmladdnormallinkfoot{VOStatistics}{http://vostat.org/} services 
(Graham et al. 2004), -- and visualization.  Composing multiple 
services into a distributed workflow architecture, 
as illustrated in Figure~\ref{fig:workflow}, with domain experts in different
areas deploying and exposing their own services, has a number of distinct 
advantages, including:
\begin{itemize}
\item Proprietary algorithms can be made available to end users without
the need to distribute the underlying software.
\item Software updates done on the server are immediately available to all 
users.
\item A particular service can be used in different ways as a component of
multiple workflows.
\item A service may be deployed close to the data source, for efficiency.
\end{itemize}
Interactive deployment and control of these distributed services will
be provided from a workflow manager. We expect to use NVO services for
data access -- images, catalogs, and spectra -- as well as the NVO
registry for service discovery.

\section{Graduated Security}

\begin{figure}[hbt] 
\plotone{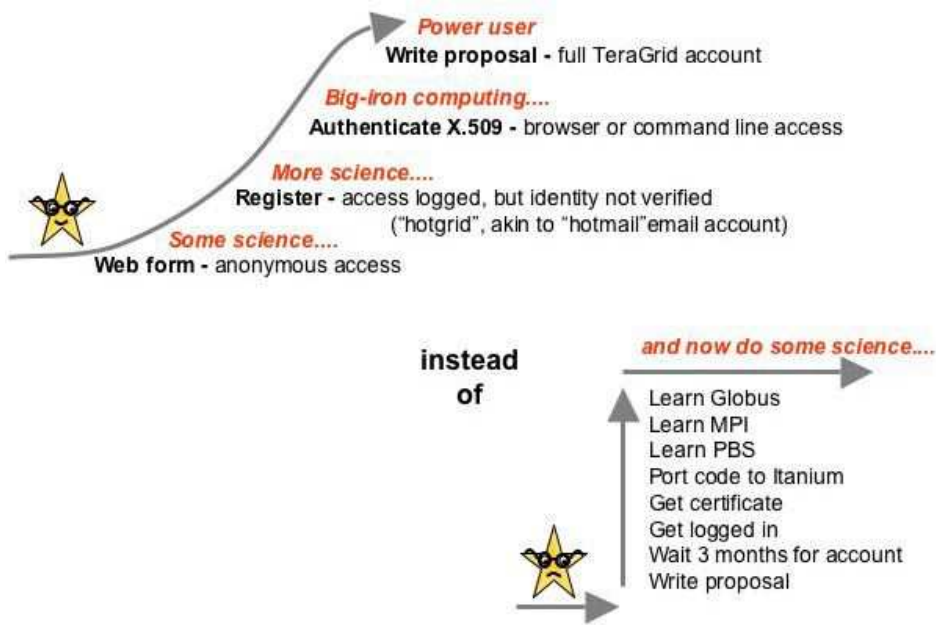}
\caption{``Graduated security'' will shorten the hurdles that stand in the way
of scientists who would like to take advantage of the power of computational 
grids for their research.}
\label{fig:grad_sec} 
\end{figure}

As described in Section~\ref{sec:workflow}, 
much of the pipeline and mining software for Grist will be built in
the form of web services. One of the reasons for building services is
to be able to use them from a thin client, such as 
a web browser. However, for such services to be able to
process private data or use high-end computing, there must be strong
authentication of the user. The VO and Grid communities are converging
around the idea of X.509 certificates as a suitable credential for
such authentication. However, most astronomers do not have such a
certificate, and we don't want to make them go through the trouble of
getting one unless it is truly necessary. Therefore, we are building
services with ``graduated security'', meaning not only that small
requests on public data are available anonymously and simply, but also
that large requests on private data can be serviced through the same
interface. However in the latter case, a certificate is
necessary. Thus the service ``proves its usefulness'' with a simple
learning curve, but requires a credential to be used at
full-strength (see illustration in Figure~\ref{fig:grad_sec}).

\section{Palomar-Quest Data Mining}

\begin{figure}[hbt] 
\plotone{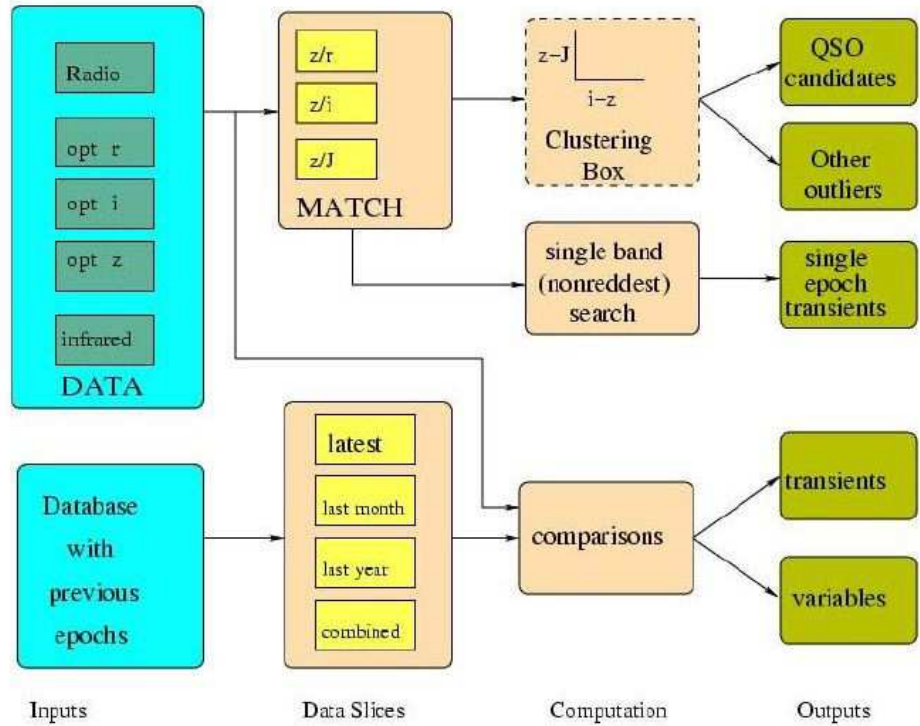}
\caption{A schematic pipeline to look for quasars, transients and
other variables. Combining multi-filter information with multi-epoch
datasets through a set of well established techniques will yield a
rich set of astronomically interesting objects.}
\label{fig:pq-pipeline} 
\end{figure}

A key science-driven workflow we are constructing is illustrated in
the schematic in Figure~\ref{fig:pq-pipeline}.  The primary objectives
are to search for high redshift quasars and optical transients in data
from the Palomar-Quest sky survey.  The pipeline begins by federating
multiwavelength datasets, and
matching objects detected with the {\em z} filter with catalogs at
other frequencies.  Cluster analysis performed on the resulting
color-color plots (e.g., {\em i-z} vs.\ {\em z-J}) yield new quasar
candidates, and outliers may indicate the presence of other objects of
interest.

Single epoch transients are indicated by objects that are detected in one 
filter but not others. An object that is detected in the
reddest filter is of special interest since it could be a
highly obscured object or a high redshift quasar.
For multi-epoch transient search,
illustrated in the lower part of Figure~\ref{fig:pq-pipeline}, we
compare new data with a database of past epochs to detect new transients
or other variable objects.

As described above, a
primary objective of the PQ survey is the fast discovery of new types of
transient sources by comparing data taken at different times. Such
transients should be immediately re-observed to get maximum scientific
impact, so we are experimenting with ``dawn processing'' on the
TeraGrid, meaning that data is streamed from the telescope to the
compute facility as it is taken (rather than days later). The pipeline
itself is being built with streaming protocols 
so that unknown transients (e.g., newly identified
variables or asteroids) can be examined within hours of
observation with a view to broadcasting an email alert to interested parties.

\section{Summary}

Grist is developing a library of interoperable grid services for
astronomical data mining on the TeraGrid, compliant with Grid and VO
data formats, standards, and protocols.  For ease of use, Grist
services are built with graduated security, requiring no more formal
authentication than is appropriate for a given level of usage.  Grist
technology is part of a Palomar-Quest data processing pipeline, under 
construction, to search for high red-shift quasars and optical transients.
More information on Grist can be found on our project web site at 
\htmladdURL{http://grist.caltech.edu/}.


\begin{references}

\reference Djorgovski, S.\ G., et al. 2004, \baas, 36, 805

\reference Graham, M.\ J., et al. 2004, \adassxiv, \paperref{P1.2.7}

\reference Mahabal, A., et al. 2004, \adassxiv, \paperref{P2.2.7}

\reference Williams, R.\ D., et al. 2004, \adassxii, \paperref{O4-3a}

\end{references}
\end{document}